# An active smartphone authentication method based on daily cyclical activity

**Chunmin Mi[1,*], Runjie Xu[1,*], Ching-Torng Lin[2] and Yadong Xiao[1]**

1   Nanjing University of Aeronautics and Astronautics, College of Economics and Management, Nanjing, 210016 China

2 Da-yeh University, Department of IT, Changhua, 5159 Taiwan

**AUTHOR CONTRIBUTIONS:** *These two authors contributed equally to this work, Both are the first authors.

**ABSTRACT**

With the increase in the use of smartphone, users must constantly worry about security and privacy. Although several biometrics or or behaviometrics active authentication systems have been developed and improved the security to a certain extent, there still exists a high risk of being descripted. Prophet algorithm is a procedure for forecasting time series data based on an additive model. This algorithm has strong robustness for missing data and trend change, and can deal with outliers well. Based on the time series behavior data of mobile terminal users, a data-driven auxiliary authentication method uses Prophet algorithm to decompose the time series of six periodic activity, namely walking, running, standing up, sitting down, lying down and jumping, and strip off the singular value, to get the inherent cycle and trend of each behavior, and to verify the legitimacy of the behavior user at the next moment. The UniMiB SHAR DATA was used to test the effectiveness of the proposed method. The experimental results show that the user simply needs to do 5 cycles of specified actions to realize the prediction of the next time series. The Error analysis of cross validation was applied to 4 different indicators, and the Mean Squared Error of the optimal result "Jumping" behaviour was only 8.20%. With these appealing features, the main contribution of this paper is to propose a smart phone user identification system based on behavioral activity cycle, which can be replicated in other behavioral studies. Another outstanding feature of such a system is the capability of fitting models using small data set by exploiting behavioral characteristics derived from periodicity and thus reducing dependence on sensor scanning frequency, therefore the system balances among energy consumption, data quantity and fitting accuracy.

## 1. INTRODUCTION

Everything has two sides, the vigorous development of smartphone industry has brought a brand new area of user-friendly applications, (such as mobile transfer, e-mail, mobile payment, smart home, etc.) [1][2], and leaves personal information more exposed to leakage risks. Research shows that attackers can detect the position of screen click on smartphones based on acceleration sensors and gyroscopes, get letters, numbers or sliding gestures on the screen, and crack the PIN password, or even extract a password from the residual oil on a smartphone screen [3][4][5]. There are also studies showing that mobile phone theft and other criminal incidents occur every year in various countries, and continue to grow. The loss of these smart devices will not only cause the consumption of property of the owners, but also cause the leakage of privacy data inside the mobile phone [6][7].

Therefore, a non-active user authentication mechanism is very necessary, which makes biometrics become a popular research field [8][9][10][11]. At present, biometric research has achieved relatively high accuracy, but it is still difficult to be applied in mobile devices, which mainly results from dependence on hardware, and high sensitivity of physiological mechanisms to some environmental factors. For example, facial recognition requires that the sensing device must be correctly aligned with the face, while speech recognition has more stringent environmental requirements for background noise. Some scholars have proposed a activity recognition model based on mobile acceleration sensor [12][13][14], which can operate normally without other hardware devices, and its authentication mode is transparent, persistent, making it much more difficult to imitate. For example, Sitova etc. studied the physical characteristics of 100 subjects who tapped or held their mobile phones while sitting or walking to achieve behavioural recognition, while also considering energy consumption [15]. They carry out surveys on the effects of activity identification (authentication, BKG, and energy consumption on friendship) from multiple perspectives. These are truly important innovations that can be commercialized and universally embraced in the future.

However, activity identification system above seems to need improvement. In the study of activity recognition on mobile phone sensor, the scene focuses on the active interactions between the user and the intelligent device, such as screen interaction, sensor interaction, and application interaction.

We consider an unusual but very important scenario that tends to be in extreme situations such as crime and theft.

(1) After the user's phone is stolen, the perpetrator may choose to leave the scene of the crime and steal data from the phone in a safe or technical environment, or to reset the phone for sale. The phone will have a process from the legitimate owner to the illegal holder. It can be effective to identify crisis behaviors during this period and to send warnings or other protection measures to users. These risks are often not done through fingerprint recognition and face recognition.

(2) Users can choose to give their mobile phones to someone for operation. For example, when you lend your phone to others for use, others transfer money in this time,and install the virus and other operations. The period also experienced from the legal owner to the illegal owner of the process. However, if the characteristics of sensitive operations and users can be identified during this period, additional protection measures can be taken.

If we can find the illegal operators in time and establish the privacy protection mechanism, it will be of great significance.

Based on the non-interactive scenarios, Mosenia etc. uses sensors for ubiquitous and ongoing authentication, describe how it can be extended to user identification and adaptive access control authorization [18]. Batchuluun etc. proposed the concept of fusion activity recognition and activity prediction, which captures objects through infrared sensors and realizes activity recognition technology with higher accuracy based on fuzzy system [19]. But this kind of research often requires the model to have higher identification ability, and such targets often have a large demand for data volume [20][21][22]. Users need to make multiple sets of actions to achieve higher authentication accuracy. The acquisition of data requires constant energy consumption of sensors, which leads to problems related to energy consumption.

In order to solve the above-mentioned authentication problems in non-interactive scenarios, and achieve the balance between energy consumption and authentication accuracy in the authentication process, this paper presents a framework design running in the backstage. The whole system includes three parts: activity model, recognition mechanism and prediction model. Through the training of UniMiB SHAR DATA set by Prophet algorithm, we empirically analyzed the activity characteristics of device carriers under six kinds of daily activities. The results show that the method can predict the confidence interval for the next moment after users make 5 periodic actions, and can achieve continuous time series prediction, and better verify the legitimacy of the logged-in users in the subsequent use process. By means of cross-validation Error analysis, the Mean Squared Error of "Jumping" behavior is obtained as 0.08, and the Mean value of the other three indicators is between 0.18 and 0.28.

Unlike other scenarios which require entering password, this system hopes to realize that not need to turn on the screen of the mobile phone, nor trigger device interaction. It only needs to be carried by the user to judge the legitimacy of the user. Furthermore, the system is designed to exploiting behavioral characteristics derived from periodicity. One thing here is to make out a macroscopic characteristic figure, and save more energy by lowering sensor scanning frequency. The other is to focus on periodic characteristics in preparation for situations that the data set are small, and in those cases, a proper fitting accuracy could solve possible issues caused by "cold start". It achieves a balance of energy consumption, data volume and accuracy.

## 2. MAIN IDEA

Smartphone sensors is a typical multi-variable time series classification problem. It uses sensor signals and extracts features from them to identify activities through classification. Multiple sensors on smartphones are used to capture the user's activity information to verify the users of smartphones.

The idea of activity recognition provided in this paper mainly comes from the user's personal habits in daily life. Since these habits cannot be directly depicted by data, it is difficult to be imitated and attacked. When a user carries a mobile phone (e.g. putting the phone in the pocket), the small disturbances of the device can be collected by the sensors of the mobile phone. It is worth noting that there are many kinds of actions in the mobile phone scenario, but some simple actions are easy to recur during the day, such as running, standing, sitting and so on. The recognition system in this paper mainly carries out data mining for these actions of frequency, so as to realize the effective judgment of user legitimacy.

After the deployment of the activity recognition system, the application needs to capture and train the user's specific activity to confirm whether the current activity data is in our recognition framework. Therefore, we use a single-class SVM classifier to recognize user's activity, which only determines what kind of activity the observed data belongs to.

Then we adopt machine learning to train the accumulated data of users, so as to predict the scope of actions at the next moment. The evaluation level of the model is updated by increasing the observation value continuously. If the current user's activity characteristics are abnormal, the system will be prompted to trigger the privacy protection mechanism. If the newly collected data set is within the reasonable confidence interval of the activity model, the user will be defaulted to be a legitimate user, and the new data will be used as the model update.

The ultimate purpose of this experiment is to extract periodic characteristics of human daily behaviors, so as to achieve a balance among energy consumption, data quantity and model accuracy.

## 3. DATA AND ALGORITHM

The few open data sets of terminal sensors can be divided into two main categories: data set acquired by environmental sensors and data set acquired by wearable devices. In this paper, considering that the data set studied must satisfy the application scenario without human-machine interaction, we choose UniMiB SHAR data set [26].

Samsung Galaxy Nexus 19250 equipped with Android 5.1.1 and Bosh BMA220 acceleration sensors is used to collect the data set. A total of 11711 activities were collected from 30 subjects aged 18 to 60. The activities were divided into 9 kinds of daily activities and 8 kinds of fall tests. The data set includes three vertical axis accelerations with a constant sampling rate of 50 Hz.

The main analysis algorithm used in this paper is Prophet [27]. Proposed by Facebook, this algorithm is a prediction algorithm based on time series decomposition and machine learning. It has superior performance in dealing with default values and large-scale automatic prediction.

Prophet uses Decomposition of Time Series, which includes the main conditions needed for time series data analysis: overall trends, periodicity and noise. Its basic model is:

$$y(t) = g(t) + s(t) + h(t) + \varepsilon_t .$$

In which, $g(t)$ indicates the growth function to fit the non-periodic changes, $s(t)$ indicates the periodic qualitative changes, $h(t)$ indicates the special changes caused by fixed time periods (such as holidays), and $\varepsilon_t$ indicates the noise.

The growth function $g(t)$ is defined as a logical function:

$$g(t) = \frac{C}{1 + exp(-k(t-b))} .$$

This function is similar to the population growth function, in which $C$ is capacity, $k$ is growth rate and $b$ is offset. With the increase of time, $g(t)$ will tend to be $C$.

However, this model alone does not satisfy the actual requirements, because the growth rate may vary over time. Therefore, a "change point" can be set to represent the time node of the change in growth rate. At this time, $t = s_j$ is a turning point.

$$a_j(t) = \begin{cases} 1, & \text{if } t \geq s_j, \\ 0, & \text{otherwise.} \end{cases}$$

Under the influence of "change point", in order to make the function continuous, the rate at time $t$ is:

$$k + a(t)^T \delta .$$

In this case, the mathematical needs the following treatment:

$$\gamma_j = \left(s_j - b - \sum_{l<j}\gamma_l\right)\left(1 - \frac{k + \sum_{l<j}\Delta_l}{k + \sum_{l\le j}\Delta_l}\right).$$

The segmented growth logic model can be obtained:

$$g(t) = \frac{C(t)}{1+\exp(-(k+a(t)^T\delta)(t-(m+a(t)^T\gamma)))}.$$

For forecasting problems that do not exhibit saturating growth, we can also use trend functions:

$$g(t) = (k+a(t)^T\delta)t + (b+a(t)^T\gamma).$$

Among them,

$$a(t) = (a_1(t),\cdots,a_s(t))^T, \delta = (\delta_1,\cdots,\delta_S)^T, \gamma = (\gamma_1,\cdots,\gamma_S)^T.$$

The classic case of periodicity in Prophet is the periodic changes brought about by seasonality, which are approximated by Fourier series.

$$s(t) = \sum_{n=-N}^{N} c_n e^{i\frac{2\pi n}{P}}.$$

In dealing with periodic problems, Prophet's task is to fit $c_n$ coefficient. The larger $N$ is, the higher the ability to fit complex cycles is.

The variable represented by $h(t)$ is very important in the analysis of time series, such as the significant changes of the number of people in shopping malls on weekends and festivals. The method Prophet uses for this purpose is to extract these dates separately and set a virtual variable.

The periodic function in the interval is expressed by Fourier series:

$$s(t) = \sum_{n=1}^{N}\left(a_n\cos\frac{2\pi nt}{P} + b_n\sin\frac{2\pi nt}{P}\right).$$

After that, the periodic term decomposed from the time series is:

$$s(t) = X(t)\beta.$$

And the initialization of "$\beta$" is:

$$\beta \sim \mathrm{Normal}(0,\sigma^2).$$

The larger the $\sigma$ here, the more pronounced the seasonal effect. On the contrary, the effect of season is less obvious.

Finally, the other items of the model are represented by indicator function:

$$h(t) = Z(t)k = \sum_{i=1}^{L} k_i.$$

Prophet uses the cyclic analysis method of manual regulation and automated prediction to make large-scale data prediction. It uses simulated historical predictions to evaluate performance outside the sample, and modifies predictions with problems. It has the following characteristics:

• It can process large-scale, fine-grained data and flexibly switch data resolution between micro and macro levels.

• Compared with other modeling tools, Prophet can be used without continuous time series data, so the problem of data defaults can be neglected.

• It can identify mutation points and abnormal values, and have a variety of anti-mutation and regulation methods.

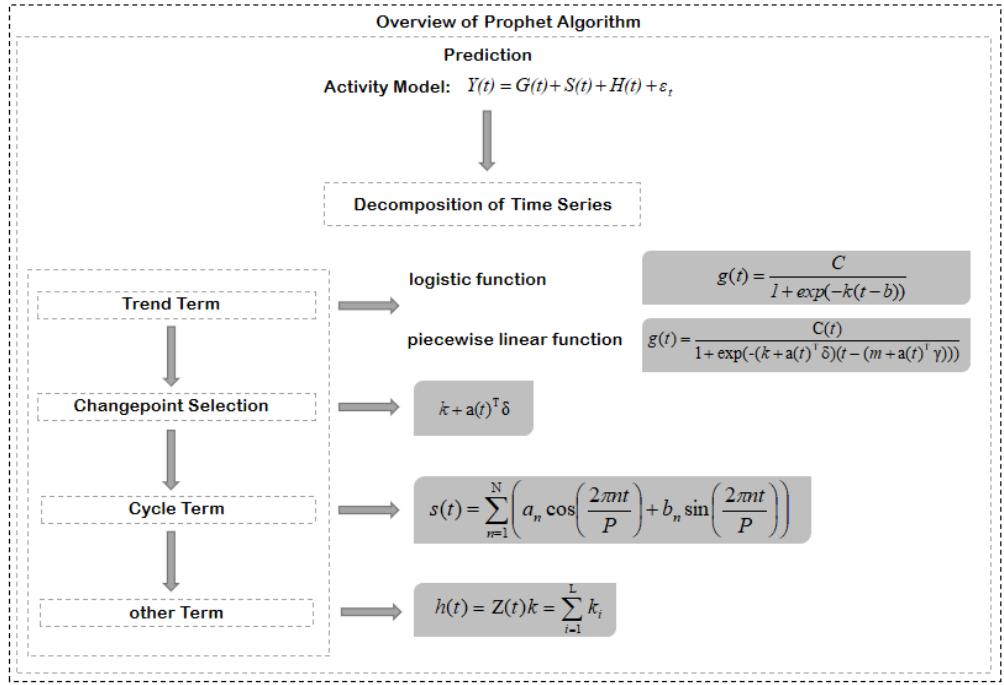

Fig.1 Overview of Prophet Algorithm.

## 4. RECOGNITION MODEL

According to the research on activity recognition in relevant literature [23][24][25], we propose a activity recognition model for smart devices in non-interactive scenarios.

Firstly, in order to accurately characterize an individual's unconscious activity habits, the activity model should include two characteristics: user activity and device response. For the simple and easy repetitive actions we have studied, there are the following basic elements: whether there is periodicity in frequent repetition of actions, the singular values caused by the accuracy of sensors in measuring, and the noise caused by other factors.

The device observes various actions in a day's record, and these actions may be repeated in a day. We regard it as a system to be expressed by $O_i \{S_i, f_{a1}, f_{b1}, f_{c1} \ldots, f_{in}\}$, in which $O_i$ represents the whole behavioural system of the user whose serial number is $i$ detected by the mobile phone sensors; $S_i$ represents a series of singular values which are inevitable when sensors record, this is because that minor changes of different gestures and different holding postures will cause difference of vibration values[23][28]; $f_{in}$ represents the $i$ type of actions being observed for the $n$ times.

For any action $i$, we express the activity model of the action as $Y(t)$:

$$Y(t) = G(t) + S(t) + H(t) + \varepsilon_t$$

In which $Y(t)$ represents the prediction model of the activity at $t$ time; $G(t)$ represents the growth function to fit the activity without obvious periodicity; $S(t)$ represents the periodicity of the activity; $H(t)$ represents the singular value of sensors; $\varepsilon_t$ represents the noise caused by other environmental factors.

## 5. BALANCE AMONG ENERGY, DATA VOLUME AND ACCURACY

In practical applications, as a safety auxiliary function for identity authentication, activity recognition often needs high credibility of conclusions, which may need a long period of model training to provide better service to users. Although continuous observation of sensors can indeed collect abundant behavioural information for the owner's personal model training, the main energy cost of behavioural recognition is caused by sensors, and continuous observation may result in useless energy consumption [24][28]. For example, when users put mobile phones on the desktop for users sitting for a long time or users playing games, it is unnecessary for the sensors to be activated continuously.

For the major smartphone brands in daily use, we checked their suppliers of triaxial acceleration sensors and listed their energy consumption.

Table.1 Triaxial acceleration sensors and associated power consumption from major smartphone brands.

| Smartphone | Sensor Chip | Output Signal |
|---|---|---|
| Galaxy Nexus 19250 | Bosch BMA220 | Supply Current in Normal Mode: 250uA<br>Supply Current in Low Power Mode: Less Then 10uA<br>Supply Current in Suspend Mode: 1uA |
| Galaxy Nexus S5 | Invensense MPU 6050 | Normal Operating Current:500uA<br>Low Power Mode:10uA-1.25Hz;20uA-5Hz; 60uA-20Hz;110uA-40Hz |
| GALAXY S6 Edge | MPU-6500 | Normal Operating Current:500uA<br>Low Power Mode:10uA-1.25Hz;20uA-5Hz; 60uA-20Hz;110uA-40Hz |
| Iphone6 plus | Bosch BMA280 | Normal Operating Current:130uA<br>Low Power Mode:6.5uA |
| Iphone 7/7 plus | BMA 253 | Normal Operating Current:14.5uA<br>Low Power Mode:6.5uA |
| Xiaomi smart bracelet | ADXL362 | Normal Operating Current:1.8uA-100Hz;3uA-400Hz<br>Suspend Mode：10nA |

However, if the sensors are shut down for a long time, the incident may still occur in the meantime, and the risk will increase with the increase of detection stop time. This reflects the difference between identity recognition and activity recognition: the purpose of activity recognition is accuracy, while that of identity recognition is safety. In order to reduce energy consumption without decreasing recognition accuracy [29][30], a good observation timing mechanism still needs to be designed.

For this purpose, we design a mechanism to determine the time of starting and shutting down sensors, so as to achieve the balance between prediction accuracy and energy consumption. The aim is to use the terminal equipment efficiently for the training of "master model", under the assumption that the device is carried by a single user on a daily basis.

To achieve as high accuracy as possible, the main point of our decision strategy is use as less sensors as possible and maintain sufficient confidence on identifying users.

First of all, we attach vital importance to personal privacy and account security, and save energy from low-risk activities that requires less protections, such as watching movies or playing single-player local games.

Another choice is regular security analysis service, combined with random examinations for higher credibility. During data collection, previous researches generally adopt low duty-cycled operations, which is switching the sensors between sleeping mode and working mode to save battery [28][32][33]. If the user's manner has changed in a way far from the data observed previously, the device would forbid some access to system kernel and start continuous fitting until the evaluated risk has dropped down to the threshold.

Compared to typical time series data (such as foreign exchange price and house price), the movement data of human bodies display a significant feature of cliff drops, and the distribution of drops and discrete data points has a recognizable structure, which could be used in pattern matching during period analysis, where fourier transform (FT) could do a better job.

Any data will go through three steps under this mechanism: identification, judgment and retraining.

• Firstly, the system needs to recognize the action of the newly collected data. In many literatures, it can be found that models such as SVM can tag the user's motion type [28].

• Then, the new collected data are compared with the predicted results of the master model. In the following case study, we give the confidence interval of activity recognition prediction results, which we define as "Tolerable Error (TE)".

• Finally, if there is this error range between the new data and the predicted results of the original model, the system will not be warned of potential safety hazards by adding the new data set to the later iteration process of the model; if the predicted results of the model are far from the actual measured values, or have exceeded the confidence interval range, the data acquisition intensity will be enhanced, and the legitimacy of users will be further confirmed.

## 6. PERFORMANCE EVALUATION

Consider a situation where an attacker can access a smartphone and already has a password (such as a PIN or fingerprint) to unlock the device. As a result, attackers often have access to sensitive information. For example, an attacker might steal application accounts and personal photos, or use those accounts to send or post fake messages on social networks for malicious purposes. The goal of our work is to develop smart phone active authentication based on motion sensors, by analyzing the behavior of sensors in users' daily use, in an unobservable way. Therefore, when the attacker interacts with the smartphone, he does not know that he is in the process of implicit identity detection, and reports the detection results to the operating system, which performs corresponding operations. It's worth noting that we primarily consider user authentication of the smartphone owner, since smartphones are typically private and are not usually shared by others.

Among UniMiB SHAR DATA sets, we selected six kinds of data, Walking, Running, Standing up, Sitting down, Lying down and Jumping, as the reference actions of activity recognition. These six kinds of actions are different from the complex behaviours such as dining and outdoor mountain climbing. The selected actions in this paper are called simple actions. From our experimental process, simple activity has higher recognition accuracy in activity recognition and identity authentication. On the one hand, these behaviours are moderately complex and very common in life, which are very suitable as reference variables for identity authentication. On the other hand, these behaviours have their own periodicity in the process of movement, making it easy to mine features.

If we observe through the three vertical acceleration data in the data set directly, we find that the periodicity of each action cannot be obtained directly, and it is difficult to define the activity recognition model of the subject. Unlike previous research methods to achieve activity recognition and user authentication, we use time series prediction method to predict the next set of time series through training data sets, and then use prediction to obtain confidence intervals to be compared with new data, so as to achieve the purpose of judging legitimate users and intruders.

In this experiment, we use Prophet algorithm to identify mutation points and abnormal values. As shown in Figure 1 below, black represents the discrete points of the original time series, and dark blue lines represent the values obtained by fitting the time series, while light blue region represents the confidence intervals of time series.

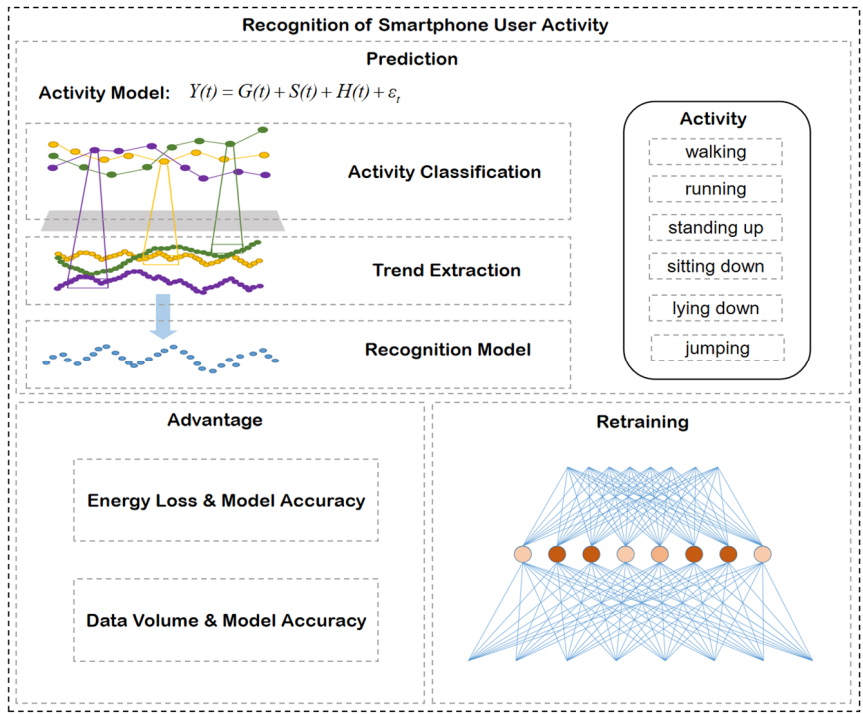

Fig.2 The system framework of Recognition of Smartphone User Activity.

## 6.1 Walking and Running

As the others, walking and running are typical periodic body movements cooperated by feet. They both triggers similar distribution pattern, which is shown in Figure 3 (black dots from triaxial acceleration sensors).

• The transferring processes back and forth between the two behaviors have a significant pattern. The running mode has a wider amplitude, making it easier to be detected when there came a violent change in the triaxial data during a steady walking. This might imply emergencies, for example, mobile phone theft cases, or the user simply has the mood to run. However, under none of these circumstances is safe to make online payments

• In the recognition and prediction of walking and running, errors are introduced from the magnitude of cycle period and movement intensity. In the real world, the period is generally short, whereas the cycling would last long. Thus consumes more energy on sensors to increase their scanning frequency, only not to lose pattern details to a halt. The discrete points in Figure 3 are already mixed together, bringing difficulties in period analysis. The industrial market has to make a choice between energy consumption and accuracy, and so far now they prefer battery life time.

• Noises in real-life situation could cause catastrophe if the judging system is not sufficiently smart

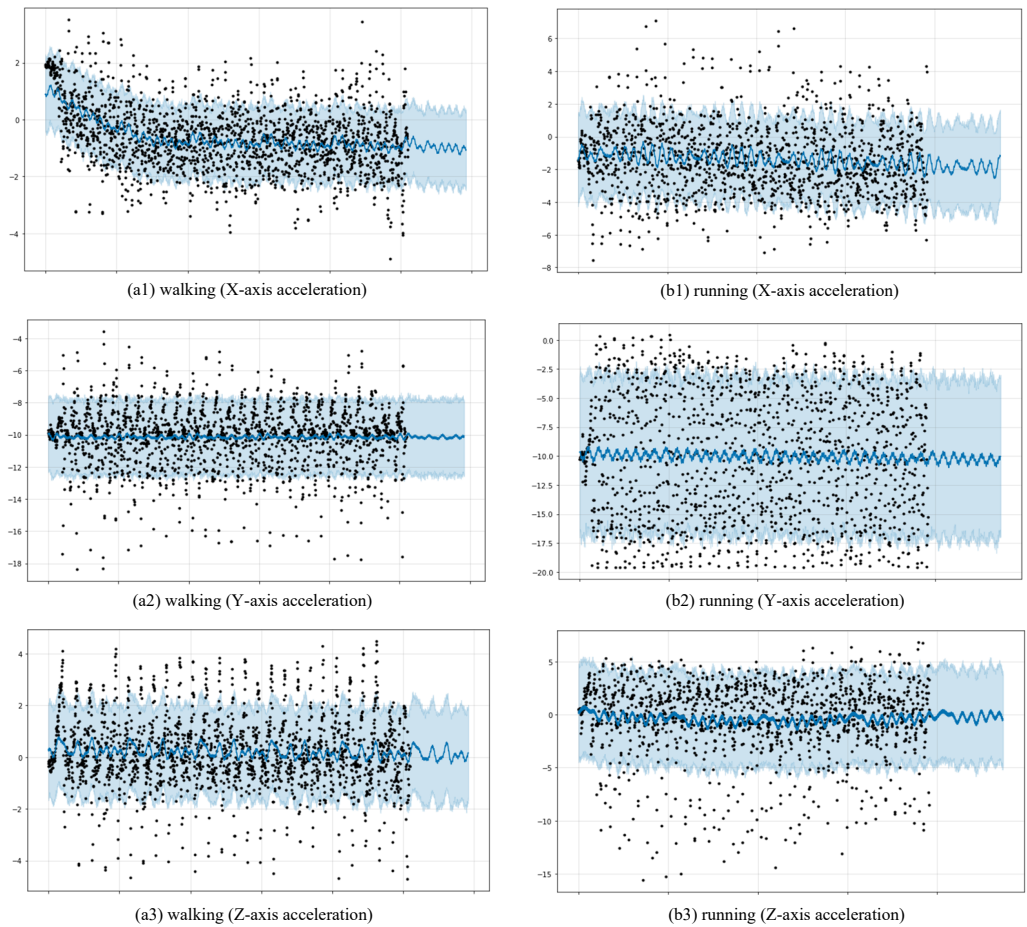

Fig.3 Experimental results of acceleration data of "walking" and "running" on X axis, Y axis and Z axis respectively.

## 6.2 Standing up、Sitting down and Lying down

Apart from walking and running, we noticed that the transformations between resting and moving, back and forth, have obvious break point and drop point structures, as in Figure 4. The cycle time is shorter so that we could trigger sensors at lower scanning frequency to extend battery life.

Take the x-axis acceleration data for example here, the maximum of three cases is stable at roughly 10 units, so is the same as other feature points, for example, the densest points. Data of y-axis and z-axis show similar properties since there is no difference between the three dimensions of a rigid body. The slopes and amplitude of acceleration curve reveal more information of user behaviour.

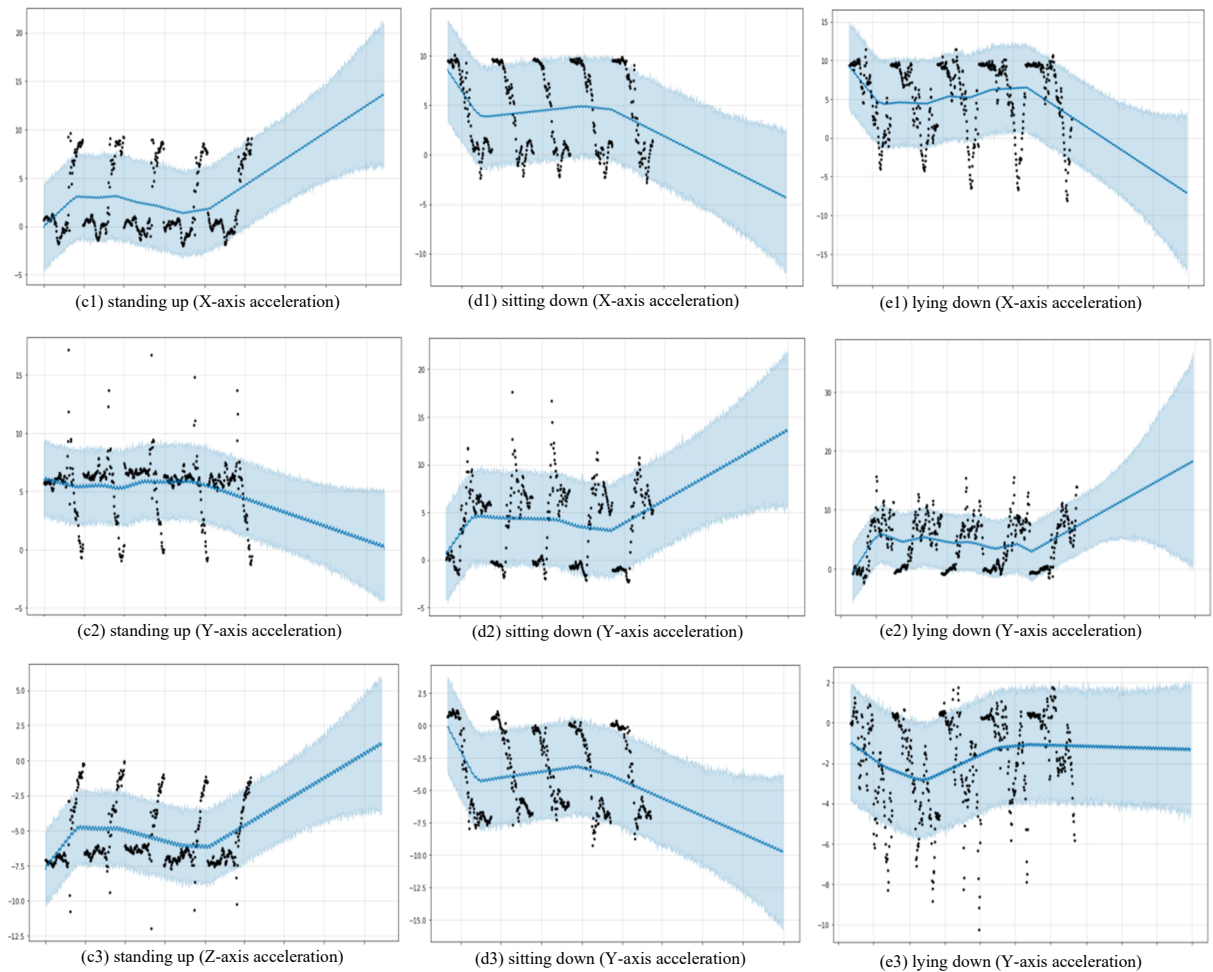

Fig.4 Experimental results of acceleration data of "standing up", "sitting down" and "lying down" on X axis, Y axis and Z axis respectively.

## 6.3 Jumping

The fitting result of jumping is the most fascinating one since it combines all the advantages we mentioned above

First, jumping is special in that some intrinsic patterns or habits of it cannot be controlled by motor nervous system. Years of movements would establish a profound cooperation manner muscles and bones, and willpower cannot cover out signatures like that. For example, in the studies of landing techniques, Van et al [35]. found that jumping with a worn knee shows a greater ankle dorsiflexion moment. Besides, Williams et al [36]. found significant changes in output power of body motor system, which implies exercises in the past and even the genes have decisive influences on jumping pattern. This reveals possibilities to take gene and environment into account during activity recognition, although the realization margin is small for now.

According to the previous literature, we hope to analyze the behavior characteristics of people through the gait trajectory[37][38].

Discrete point series in Figure 5 implies the time taken by taking off, landing and buffering is almost equal between different jumps. The distribution of acceleration peaks and valleys is similar, too. The confidential intervals colored black blue and light blue are the fitting results of Prophet. Those properties reflect personal movement patterns. It's of vital importance to control the upper and lower interval range, which is also noticed in y-axis and z-axis data. Prophet is smart and flexible enough to judge the true situation despite singularity points.

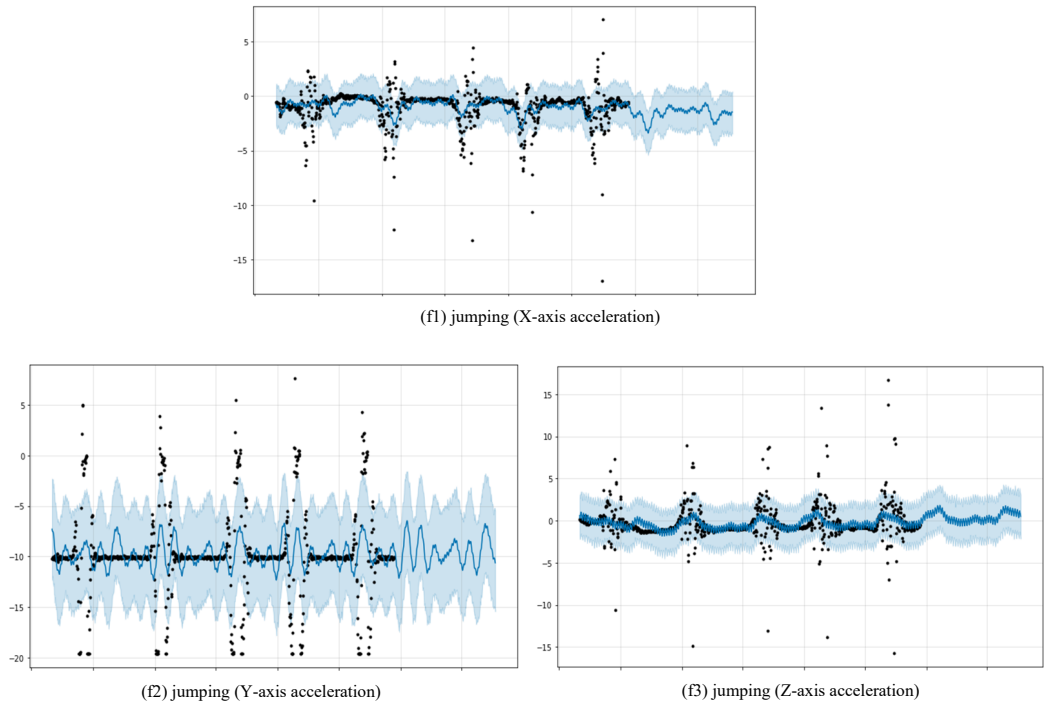

(f1) jumping (X-axis acceleration)

(f2) jumping (Y-axis acceleration)　　　　　　　　(f3) jumping (Z-axis acceleration)

Fig.5 Experimental results of acceleration data of "jumping" on X axis, Y axis and Z axis respectively

　　According to the results, Prophet makes a prediction of 300 time nodes in the future. The X-axis, Y-axis and Z-axis values of the subject moved with equal amplitude around the midline respectively. The magnitude of the amplitude indicates the subject's behavioural habits in daily activities. Because of the intenseness of activities, there are singular values in the data, but the singular values are often distributed in a fixed area, which is captured by the anti-mutation ability of Prophet algorithm, so it also has a certain trend prediction in its future prediction.

　　In order to test the ability of Prophet algorithm in recognizing human activity and realizing identity authentication, we cross-validate the experiment.

　It is worth noting that indicators such as FAR (false acceptance rate), FRR,(false rejection rate) and Receiver Operator characteristic Curve are not used in the effectiveness monitoring of the model in this paper.Although these indicators are widely used in fingerprint recognition, face recognition and other aspects.

　　　There are several reasons for this:
(1) The features of face recognition and fingerprint recognition are fixed. This means that each person's features and fingerprints are unique. But in our test scenario, there is too much randomness in human behavior, even in a repetitive action. When running, for example, people may raise their hands to dry their foreheads, affecting the frequency at which movement is detected. At this time, RAR, FRR and ROC indicators were used as the basis of monitoring effect, so it was obvious that prophet could not play its advantages.
(2) As the Prophet algorithm provides a confidence interval of time series represented by a light blue region while fitting the prediction, it can be applied in various scenarios full of randomness. However, this confidence interval can be adjusted according to the requirements of the experimenter, that is, the monitoring accuracy can be changed. At this point, FAR, FRR, ROC and other indicators are used on prophet, and the final error will be large, because prophet can increase the confidence interval to make the evaluation results of these indicators higher.
(3) Since the data we used only contained 5 repeated cycles per action, we could not achieve very high RAR, FRR and ROC. But our innovation is to use these finite action cycles to characterize what a person's behavior looks like on the X-axis, Y-axis, and z-axis. Our goal is not to measure accuracy and accuracy, but to aid other security measures that characterize user behavior cycles without the user being aware of them. Our framework doesn't provide the same security multiple times a day as face recognition and fingerprint recognition. It's more like a final layer of firewall.

The first 500 time series of Jumping data are taken as training set, and then the prediction is made every 100 groups, and the real data are used to test the prediction data. A total of five groups of cross-validation are performed, as shown in Figure 6. The results show that after the user makes the second cycle of actions, Prophet algorithm can predict the user's activity habits well. The Prophet algorithm is effective in predicting, modelling and identifying human activity data, and can be used as an auxiliary authentication method for user safety.

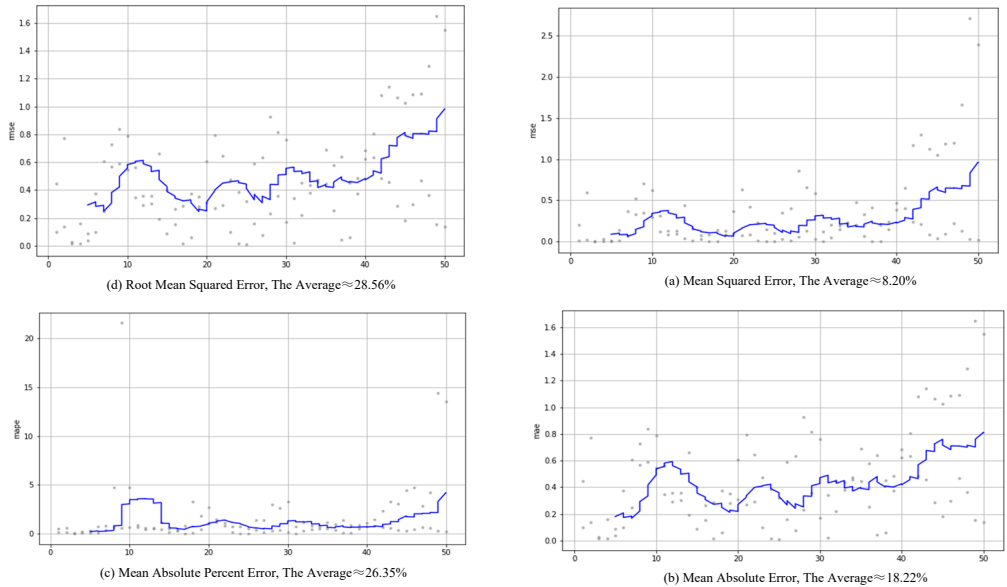

Fig.6 Prediction Error in Cross-validation of Time Series

## 7. CONCLUSION

In this experiment, six groups of common behaviours in daily life are used as model training data for human activity recognition. After using Prophet algorithm to model and predict six groups of activity data, the experiment has achieved good results and proven that the activity recognition and identity authentication system in non-human-machine interaction scenarios can be realized.

In real-life scenarios, we consider that the user's own activity habits are not unchanged. If a fixed model is always used to predict and identify the user's activity habits, it would not be accurate enough. Only the continuous change of the model can better adapt to the owner's own activity habits. The Prophet algorithm is very adaptable to periodic data and can be adjusted gradually according to the later changes, which is a step forward for activity recognition. In addition, Prophet algorithm is very suitable for time series data with missing values, which means that the shutdown and placement of mobile phones will not cause the failure of the model, promising more reliable solutions of the identification based on activity data.

In many scenarios, the application of biometric authentication assisted by activity data has broad prospects. We start from periodic activity characteristics to found a balance point among energy consumption, data quantity and fitting accuracy. Only after the owner of the device is required to carry the device for a period of time, the data accumulated by the user for 5 cycles of actions is sufficient to describe a recognition framework for the device. Special procedures catering for small data set would provide a solution for "cold start" problem. If an attacker maliciously cracks a physical password or uses a mobile phone to perform sensitive operations (such as transferring, trading, checking bank card information, etc.), activity model can be used to continuously and invisibly authenticate the attacker's identity and prevent the risk of information leakage.

**Acknowledgments:** This work is supported by the National Social Science Foundation of China (17 BGL055) and E-commerce innovation project fund of NUAA (2019EC01). We acknowledge Runyu Meng at University of Science and Technology of China (USTC) for providing inspiring advice in researching method and paper writing.